\documentclass[conference]{IEEEtran}
\IEEEoverridecommandlockouts
\usepackage{cite}
\usepackage{amsmath,amssymb,amsfonts}
\usepackage{algorithmic}
\usepackage{graphicx}
\usepackage{subfigure}
\usepackage{textcomp}
\usepackage{xcolor}
\usepackage{multirow}
\usepackage{array}
\usepackage[section]{placeins}

\def\BibTeX{{\rm B\kern-.05em{\sc i\kern-.025em b}\kern-.08em
    T\kern-.1667em\lower.7ex\hbox{E}\kern-.125emX}}
\begin{document}

\title{Reconfigurable Intelligent Surface MIMO Simulation using Quasi Deterministic Radio Channel Model
\thanks{The authors are with Huawei Technologies Co., Ltd. Their e-mails are \{semyon.dorokhin, aderkina.anastasia, Vladimir.Lyashev\}\{@huawei.com\} and lysov.pavel@huawei-partners.com}
}

\author{\IEEEauthorblockN{Semyon Dorokhin, Pavel Lysov, Anastasia Aderkina, Vladimir Lyashev}
}

\maketitle
\begin{abstract}
Reconfigurable Intelligent Surface (RIS) is a planar array that can control reflection and thus can implement the concept of partially controllable propagation environment.
RIS received a lot of attention from industry and academia, but the majority of the researchers who study RIS-assisted systems use simple Rician model. Though it is suitable for theoretical analysis, stochastic Non Line-of-Sight (NLoS) component in Rician model does not account for the geometry of deployment. Furthermore, Rician model is not eligible to evaluate 3GPP standardization proposals.
In this article we adapt the popular Quasi Deterministic Radio channel Generator (QuaDRiGa) for RIS-assisted systems and compare it against Rician model. The comparison shows that geometry-inconsistent NLoS Rician modeling results in higher estimated achievable rate.
Our method, in contrast, inherits the advantages of QuaDRiGa: spatial consistency of Large Scale Fading, User Equipment mobility support as well as consistency between Large Scale and Small Scale Fading. 
Moreover, QuaDRiGa comes with calibrated scenario parameters that ensure 3GPP compatibility.
Finally, the proposed method can be applied to any model or software originally designed for conventional MIMO, so every researcher can use it to build a simulation platform for RIS-assisted systems.
\end{abstract}

\begin{IEEEkeywords}
Reconfigurable Intelligent Surface, Intelligent Reflective Surface, MIMO, QuaDRiGa, Channel model, Capacity
\end{IEEEkeywords}

\section{Introduction}

Reconfigurable intelligent surface (RIS) is a passive planar array that can control the reflection and either perform beam-steering and beam-focusing or split
the incident beam into multiple reflected beams \cite{yurduseven2020intelligent}.
Such surfaces have a potential for capacity improvement and coverage extension \cite{liu2021reconfigurable} as a result of partial control of propagation environment.

The RIS concept quickly gained popularity, there are plenty of papers on channel estimation, joint beamforming and interference mitigation in RIS-assisted systems \cite{zheng2022survey}.
However, the majority of the researchers use Rician model for the analysis and this can greatly delay the standardization of 
RIS-assisted systems for several reasons. 
Firstly, the Non Line-Of-Sight (NLoS) component in Rician model is modeled as random, which contradicts with the deterministic nature of RIS deployment. NLoS component in Rician model does not account for the scenario geometry, which may become crucial in distributed RIS-assisted systems.
Secondly, this model does not comply with 3GPP standard \cite{3GPP2022}
and it is unclear how to calibrate it to match field tests.

One of the possible techniques for geometry-consistent simulations is Ray Tracing (RT). Though 3GPP standard \cite{3GPP2022} includes RT-based model as an alternative, such simulations need detailed maps of the environment to yield accurate results. For Macro-cell simulations it will require large number of digital maps, which is impractically difficult.

Geometry-Based Stochastic Models (GBSMs) offer a tradeoff between Ray-tracing and Rician models. 
Instead of using digital map, GBSMs generate clusters of scatterers according to calibrated distribution and only then use Ray Tracing approach. Large Scale and Small Scale fading parameters can also be included.
Several authors suggested GBSMs for RIS-assisted systems \cite{Dang2021, Jiang2021, Sun2021, basar2021indoor}, but these models have serious limitations on frequency range, number of elements, mobility of terminals, scenarios etc., as we demonstrate in section \ref{sec:Models}.
Moreover, most of them are not 3GPP-compatible and only one of them is freely accessible online.
This motivated us to build a new GBSM simulation method for RIS-assisted MIMO systems using one of the state-of-the-art conventional MIMO models.
Based on the comparison that we present in section \ref{sec:Simulators}, we decided to use QuaDRiGa  simulation platform \cite{quad2021manual, jaeckel2014quadriga}. 
QuaDRiGa is free software that supports user mobility, features spatial consistency of fading parameters and offers a wide range of 3GPP-compliant scenarios.

The main contributions of our paper are as follows:
\begin{itemize}
\item in section \ref{sec:QuaRIS} we propose a new simulation method for RIS-assisted MIMO systems based on QuaDRiGa; 
\item in section \ref{sec:Results} we compare the achievable rate of a system simulated with Rician and the proposed models and explain the huge difference.
\end{itemize}
Finally, in section \ref{sec:Conclusion} we summarize our research findings.

\section{Models for RIS-assisted systems}
Since the RIS concept is relatively young, there is still no standardized model for RIS-assisted systems. The articles on RIS use different channel models with different parameters, as we show in section \ref{sec:Models}. At the same time, the models for conventional MIMO systems are well-developed and can be adapted to RIS-assisted systems. In section \ref{sec:Simulators} we choose the best one for this task.

\subsection{RIS-assisted system models}
\label{sec:Models}
Let us consider a downlink RIS MIMO system with $N_{RX}$ antennas at the User Equipment (UE), $N_{TX}$ antennas at the Base Station (BS) and $N_{RIS}$ elements at the RIS. All RIS-assisted channel models express the total channel 
$\mathbf{H} \in \mathbb{C}^{N_{RX} \times N_{TX}}$ using the BS-UE channel 
$\mathbf{H}_0 \in \mathbb{C}^{N_{RX}\times N_{TX}}$, 
BS-RIS channel $\mathbf{H}_A \in \mathbb{C}^{N_{RIS} \times N_{TX}}$ and RIS-UE channel
$\mathbf{H}_B\in \mathbb{C}^{N_{RX}\times N_{RIS}}$:
\begin{equation}
\mathbf{H} = \mathbf{H}_0 + \mathbf{H}_B \mathbf{Q} \mathbf{H}_A,
\label{eq:basic_model}
\end{equation}
where $\mathbf{Q} = diag(\mathbf{\mu}) \in \mathbb{C}^{N_{RIS} \times N_{RIS}}$ is the RIS control matrix with diagonal elements 
$|\mu_i| = 1 \ \forall i = 1, \dots N_{RIS}$. Next, some model should be selected for $\mathbf{H}_0$, $\mathbf{H}_A$ and $\mathbf{H}_B$ channels.

Simple Freespace model can be used for theoretical estimations, e.g. to
compare RIS against Decode-and-Forward relays
\cite{de2022intelligent}.
However, this model lacks Non Line-Of-Sight (NLoS) components and therefore it is inappropriate for realistic scenarios.
To account for the multipath components, the majority of researchers \cite{zheng2022survey} adopt Rician channel model, splitting every channel matrix $\mathbf{H}_i$ in \eqref{eq:basic_model} into LoS and NLoS components:
\begin{equation}
\mathbf{H}_i = \beta \left( \sqrt{\frac{K}{K+1}} \mathbf{H}_i^{LoS} + \sqrt{\frac{1}{K+1}}\mathbf{H}_i^{NLoS} \right),
\label{eq:Rician}
\end{equation}
where $\beta$ is the pathloss, $K$ is the Rician factor, $\mathbf{H}_i^{LoS}$ is calculated according to the freespace model and $\mathbf{H}_i^{NLoS}$ is random.
For example, in \cite{bjornson2020rayleigh} the authors took into account spatial correlation by taking columns of $\mathbf{H}_i^{NLoS}$ from $\mathcal{CN}(\mathbf{0}, \mathbf{R})$ 
with correlation matrix $\mathbf{R}$ determined by $sinc$ function.

Nevertheless, such model of NLoS component is unlikely to give realistic results. Firstly, during the derivation of $\mathbf{R}$ authors in \cite{bjornson2020rayleigh} assumed that the scatterers are distributed uniformly in front of the the RIS, which hardly happens in real life. 
Secondly, the RIS position is deterministic, therefore modeling
the channels as random matrices yields an uncalibrated tool. It can be used to study properties of algorithms, but it is unclear how to tune it to match real-world measurements.

To bring RIS-assisted systems closer to standardization, researchers should use 
geometry-consistent channel models.
Good results can be achieved with Ray Tracing method, especially for indoor scenarios that allow creating highly-accurate digital maps \cite{ZhengqingYun2015}.
However, in outdoor scenarios the environment can vary a lot, especially  between urban and rural environments. Generating a sufficiently detailed map for every 
specific cell is impractically difficult. 

\begin{figure}[t]
\centerline{\includegraphics[width=\linewidth]{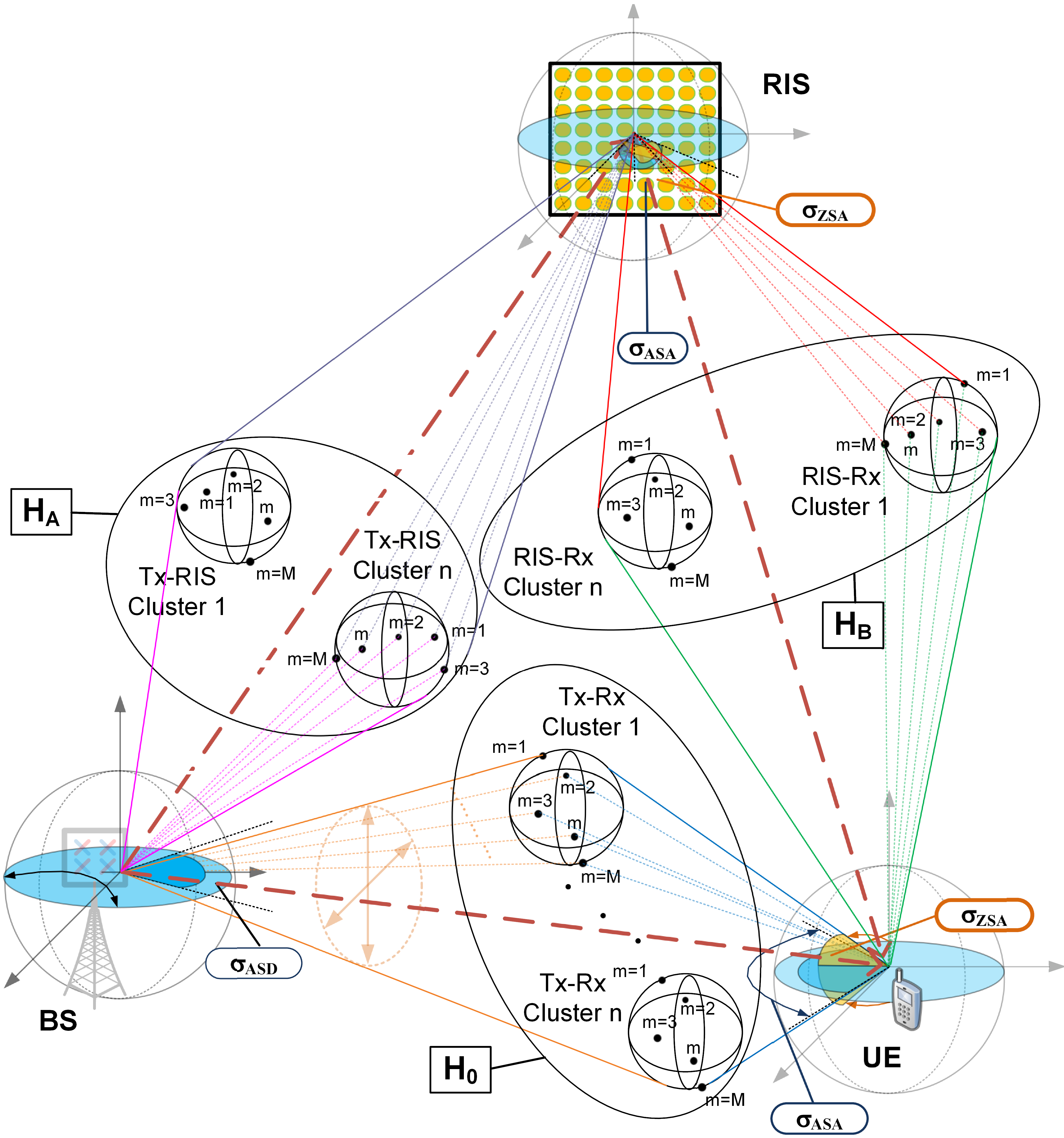}}
\caption{General structure of GBSM model for RIS-assisted systems.}
\label{fig:chan_model}
\end{figure}

Geometry-Based Stochastic Models (GBSM) offer a trade-off between the fully-determenistic Ray Tracing and almost stochastic Rician methods. 
GBSMs do not need the exact map of the environment, as the position of scaterrer clusters is generated randomly according to calibrated distributions with specific bodily angles $\sigma$, as in Fig.~\ref{fig:chan_model}. After the cluster generation, Ray Tracing can be applied. Finally, Large and Small Scale fading parameters \cite{3GPP2022} can be introduced.
GBSM models offer calibration for every typical scenario and at the same time are more accurate than stochastic ones.

Several researchers suggested different GBSMs for RIS-assisted communication.
For example, the model introduced in \cite{Jiang2021} considers only LoS part of $\mathbf{H}_A$ and $\mathbf{H}_B$ channels and NLoS part of $\mathbf{H}_0$ channel, while the model in \cite{Dang2021} features cluster generation.
However, \cite{Dang2021} does not take into account user mobility and \cite{Jiang2021} considers only one cluster in the NLoS part.
Moreover, they are limited to narrow bands and support only omnidirectional antennas.

The GBSM model from \cite{Sun2021} supports Tx and Rx mobility as well. In addition, it includes Shadow Factor and cluster evolution with death probability. Nevertheless, it calculates 
the cascaded channel $\mathbf{H}_B \mathbf{Q} \mathbf{H}_A$ pathloss based on the optimal RIS coefficients. These coefficients are derived for freespace model and may not be optimal in general. Thus, the pathloss expression may be incorrect if $\mathbf{H}_B$ or $\mathbf{H}_A$ are NLoS.

The GBSM model introduced in \cite{basar2021indoor} features pathloss for Urban Microcellular and Indoor Hotspot scenarios from 3GPP standard \cite{3GPP2022} and is available online as simulation software \cite{Basar2020SimRISCS}.
However, it supports neither Tx/Rx mobility nor wideband simulations and is implemented only for single-antenna case. 
Moreover, only LoS component is supported in $\mathbf{H}_B$ in indoor scenario and the number of  clusters is the same as for mm-Wave band, limiting simulation frequency.

Although there are several GBSMs for RIS-assisted systems, they all have rather strict constraints related either to the number of antenna elements, radiation patterns, pathloss expression, frequency range or channel conditions.
These constraints motivated us to use one of the available MIMO models and to adapt it to RIS-assisted MIMO systems.

\subsection{MIMO GBSM models}
\label{sec:Simulators}

Among all 5G GBSM simulation platforms and models, there are three that received significant popularity \cite{pang2022investigation}, namely, 
the More General 5G model (MG5G) \cite{wu2017general},
the NYUSIM channel simulator \cite{sun2017novel} and
the Quasi Determenistic Radio Channel Generator (QuaDRiGa) \cite{jaeckel2014quadriga}.

\begin{table}[t!]
\centering
\caption{State of the art MIMO GBSM models}
\label{tab:comparisonSim}
\setlength\tabcolsep{5pt}
\begin{tabular}{| m{0.06\textwidth} | m{0.08\textwidth} | m{0.05\textwidth} | m{0.07\textwidth} | m{0.05\textwidth} | m{0.05\textwidth} |}

\hline
Simulator or Model & Calibration according to 3GPP model & Massive MIMO support & LOS/NLOS transition & Moving clusters & Custom antenna patterns \\ \hline

{QuaDRiGa} & Yes & Yes &  Yes & No  & Yes \\ \hline
{NYUSIM} & No  & No & Yes &  No  & No  \\ \hline                                                             
{MG5G} & No  & Yes & No &  Yes  & Yes \\ \hline                                                             
\end{tabular}
\end{table} 

The main advantage of the MG5G model is its detailed cluster evolution.
In addition to 'newborn' and 'disappearance' cluster states, MG5G features 'survival'
state. The parameters of survived clusters are regenerated based on the new geometry and the channel coefficients are recalculated for every time interval. However, MG5G model does not support transition between 
LoS and NLoS scenarios.

NYUSIM focuses on mm-Wave range and features a channel model similar to the 3GPP \cite{3GPP2022} model.
However, the NYUSIM platform has restrictions on the number of antenna elements: no more than 128 Tx antennas and no more than 64 RX antennas. Such restrictions do not allow modeling a sufficiently large RIS to provide reasonable gain. Moreover, NYUSIM lacks custom antenna pattern support, which is crucial for physically-consistent simulations.

Compared to these two simulators, QuaDRiGa platform has no limit on the number of antenna elements and supports transition between LoS and NLoS scenarios.
Apart from that, the main advantage of QuaDRiGa is that it offers predefined channel models with parameters calibrated according to 3GPP standard channel model \cite{3GPP2022}. Different simulation scenarios can be loaded both for sub-6GHz and mm-Wave ranges.

We choose QuaDRiGa for our RIS-assisted MIMO simulations based on the comparison summarized in Table \ref{tab:comparisonSim}.
Compared to MG5G, it supports LoS/NLoS transition and in comparison to NYUSIM, it has no limit on antenna array size and supports custom antenna patterns. 
Finally, the main advantage of QuaDRiGa is that it comes with 3GPP-calibrated model configurations unlike the two other competitors.
Until the goal of RIS standardization is accomplished, using QuaDRiGa for simulations is a natural step towards it.

\section{RIS-assisted MIMO with QuaDRiGa}
\label{sec:QuaRIS}

\begin{figure}[t]
\centerline{\includegraphics[width=\linewidth]{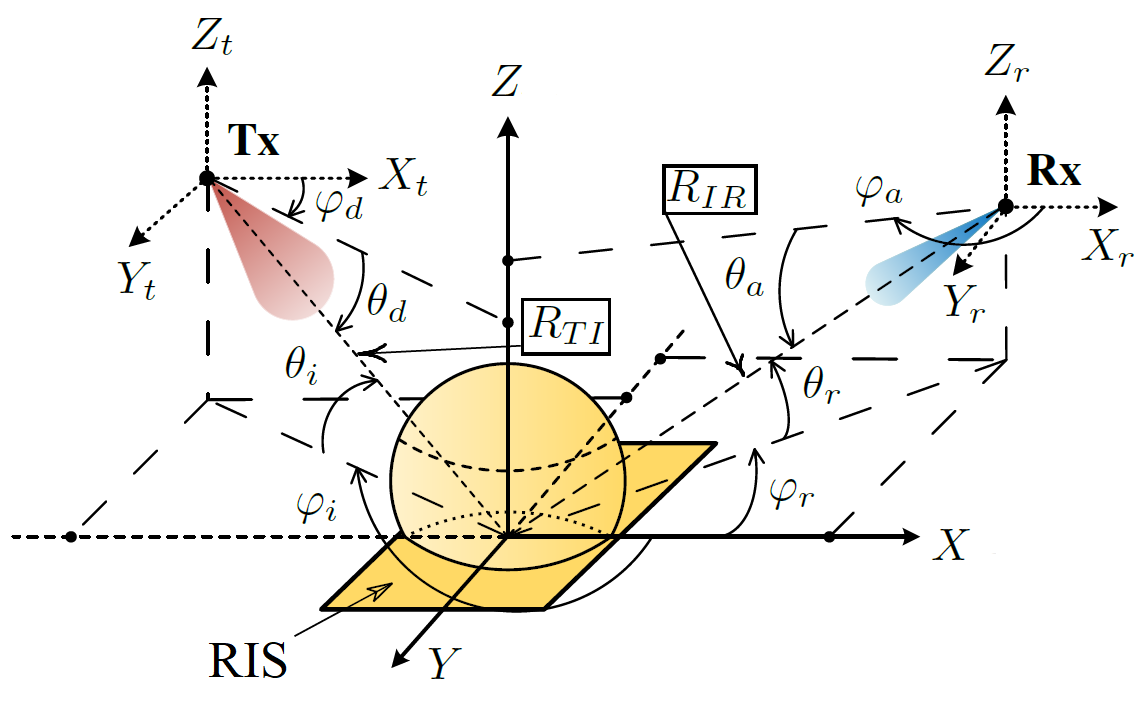}}
\caption{The geometry of the RIS-assisted system.}
\label{fig:scat_pat}
\end{figure}

In this section we propose the novel RIS-assisted MIMO simulation method that is based on the well-known QuaDRiGa platform \cite{jaeckel2014quadriga} .
The main advantage of the proposed method is that QuaDRiGa is compatible \cite{quad2021manual} with 3GPP TR 38.901 v16.1.0 and 3GPP TR 36.873 v12.5.0 standards, as well as with the mmMAGIC channel model.
Moreover, QuaDRiGa features spatial consistency of Large Scale Fading parameters and consistency between Large Scale and Small Scale parameters. In addition to this, it supports mobility of UE, which means that the proposed simulation method can be used for RIS-assisted MIMO systems with moving UEs or RIS. 

The QuaDRiGa simulation platform was originally designed for MIMO systems, so of-the-shelf QuaDRiGa does not support simulations with RIS. 
To extend it to RIS-assisted communication we model every channel in \eqref{eq:basic_model} as
conventional MIMO channel and use QuaDRiGa to obtain the corresponding channel matrix.

More specifically, we obtain the direct channel $\mathbf{H}_0$ modeling MIMO system BS-UE in a usual way.
To get the BS-RIS channel $\mathbf{H}_A$, we represent the RIS as a virtual receiver 
and simulate the BS-RIS subsystem as conventional MIMO system.
Next, we represent the RIS as a virtual transmitter and obtain the RIS-UE channel 
$\mathbf{H}_B$. 
Thus in our QuaDRiGa-based simulation platform the RIS is represented by a virtual receiver and a virtual transmitter that have the same coordinates and the same element array. At the same time, we omit the simulation between the virtual transmitter and receiver by setting the corresponding pairing parameter in configuration \cite{quad2021manual}.

Apart from 3GPP-compliant calibrated model, another detail that ensures physically-consistent simulation is the scattering pattern of RIS element. In this section we demonstrate how it can be included in the proposed simulation method.

First, let us explain the concept of scattering pattern using a simple single-antenna LoS example with single-element RIS. 
In this case, it is possible to express the received power in the cascaded subchannel using the radar equation approach:

\begin{equation} 
P_{RX} = \frac{P_{TX}G_{TX}(\varphi_d, \theta_d)G_{RX}(\varphi_a, \theta_a)\sigma_{RIS}\lambda^2}{{4\pi}^3R_{TI}^2R_{IR}^2},
\label{eq:Radar}
\end{equation}
where $P_{TX}$ is the transmitter power, $\sigma_{RIS}$ is the RIS element radar cross-section, $G_{TX}(\varphi_d, \theta_d)$ and $G_{RX}(\varphi_a, \theta_a)$ are the transmitter and receiver antenna patterns, respectively, $\varphi_d, \theta_d$ are the transmitter's Angles of Departure and 
$\varphi_a, \theta_a$ are the receiver's Angles of Arrival.
We denote the carrier wavelength as $\lambda$, and the distances from Tx to RIS and from RIS to Rx as $R_{TI}$ and $R_{IR}$ respectively, as it is shown in Fig.~\ref{fig:scat_pat}. 

\begin{figure*}[h!]
\centering     
\subfigure[CDF of $\mathbf{H}_A$ and $\mathbf{H}_B$ eigenvalues. Solid lines for $\mathbf{H}_A$, dashed lines for $\mathbf{H}_B$.]{\label{fig:a}\includegraphics[width=59mm]{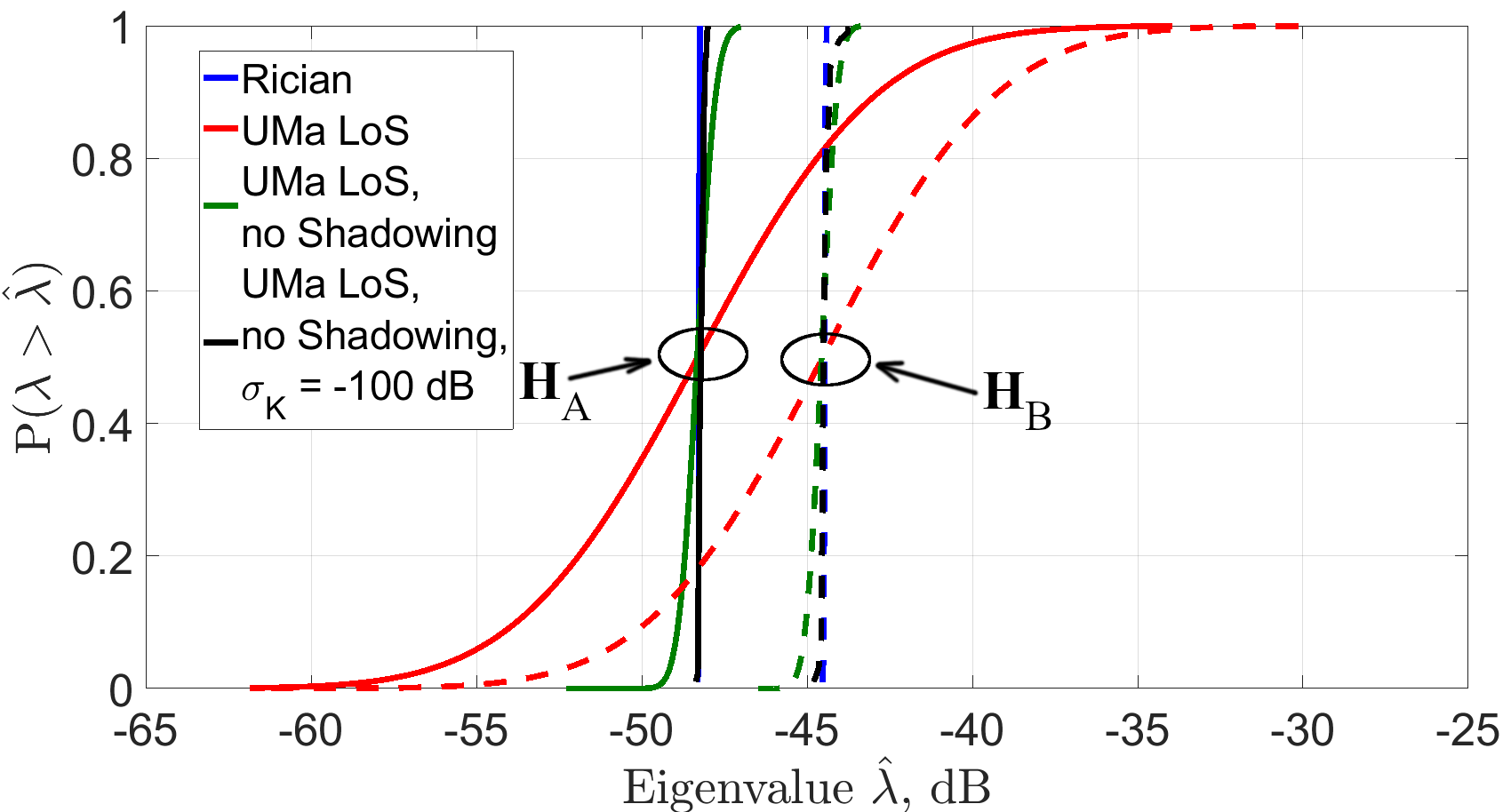}}
\subfigure[CDF of $\mathbf{H}_0$ channel. Solid lines for first eigenvalues, dashed for the second.]{\label{fig:b}\includegraphics[width=59mm]{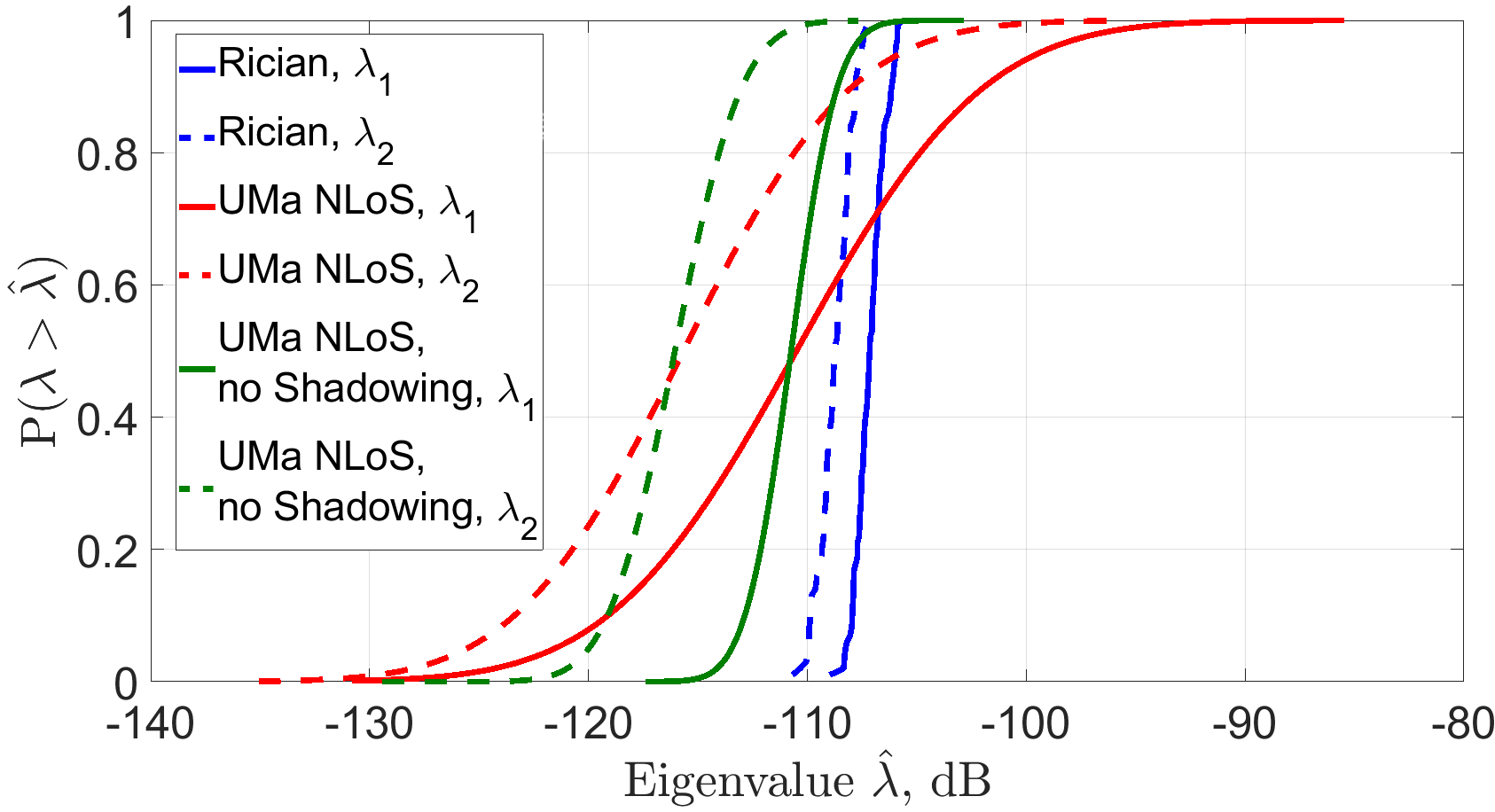}}
\subfigure[Achievable Rate CDF with (dashed) and without (solid) RIS, RIS control from \cite{zhou2020joint}.]{\label{fig:c}\includegraphics[width=59mm]{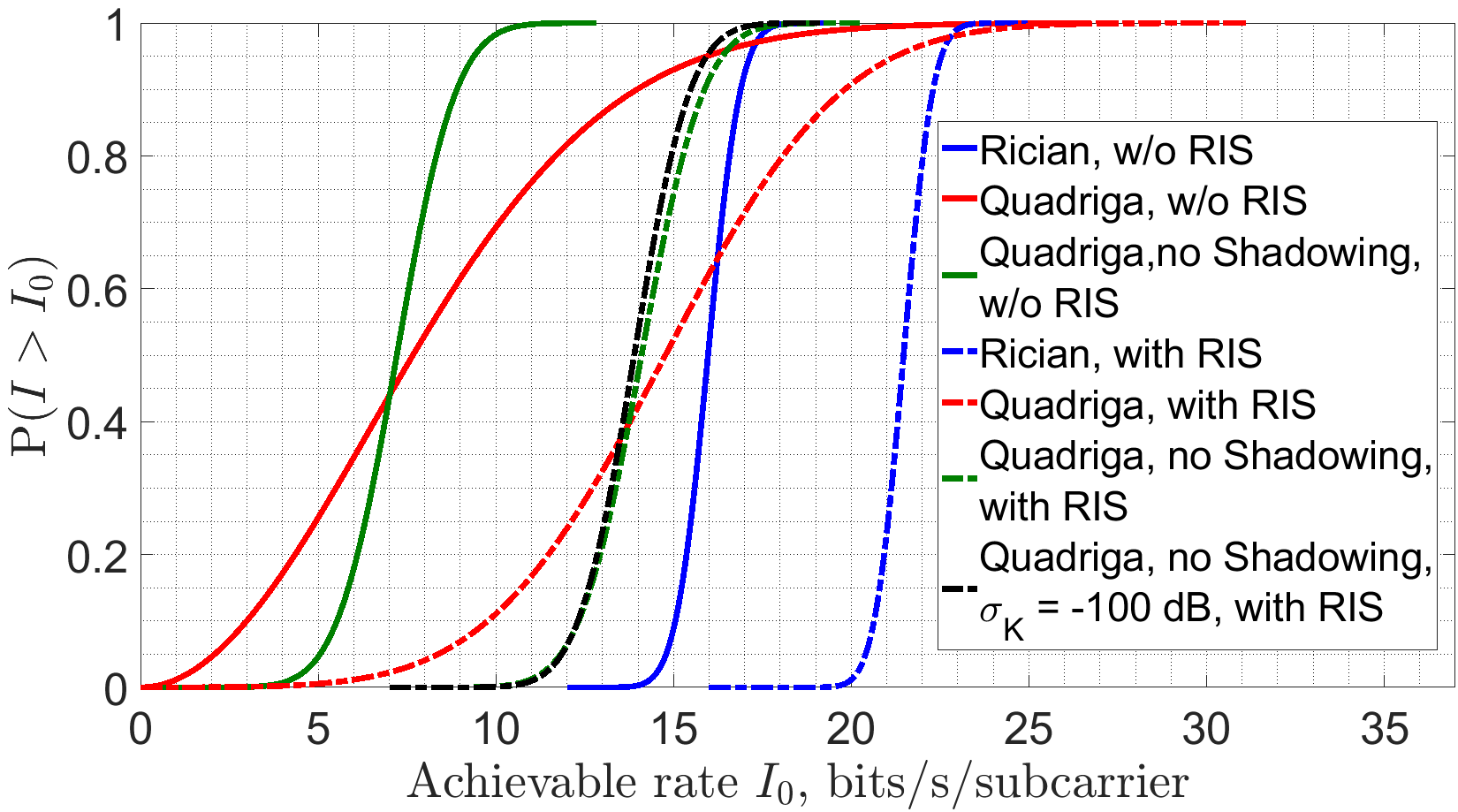}}
\caption{Comparison between the Rician and the proposed model.}
\end{figure*}

The antenna patterns in \eqref{eq:Radar} can be expressed in normalized form: $G_{TX} = G_{TX}^{max}F_{TX}(\varphi_d, \theta_d)$ and $G_{RX} = G_{RX}^{max} F(\varphi_a, \theta_a)$, as in \cite{9206044}. Supposing one RIS element can be represented as a rectangular plate of conductive material of size $a\times b$, we can obtain similar expression using eq. (11.44) from \cite{ConstantineBalanis2016}:
\begin{equation}\label{eq:RCS3DGen}
	\sigma_{RIS}(\varphi_i, \varphi_r, \theta_i, \theta_r) = 4\pi\left(\frac{ab}{\lambda}\right)^2F(\varphi_i, \varphi_r, \theta_i, \theta_r),
\end{equation}
where $\varphi_i, \theta_i$ are azimuth and elevation angles of incidence, $\varphi_r, \theta_r$ - corresponding angles of reflection, as in Fig.~\ref{fig:scat_pat}.
$F(\varphi_i, \varphi_r, \theta_i, \theta_r)$ is the isolated element scattering pattern, a normalized function that takes values between 0 and 1 and describes angular properties of scattered field.

As it follows from macroscopic reradiation model introduced in \cite{degli2022reradiation}, the scattering pattern of RIS element can be factorized as:
\begin{equation}
F(\varphi_i, \varphi_r, \theta_i, \theta_r) = 
F(\varphi_i, \theta_i) F(\varphi_r, \theta_r),
\label{eq:F_decomp}
\end{equation}
where $F(\varphi, \theta)$ is the antenna pattern of RIS element.
Decomposition \eqref{eq:F_decomp} helps us to model the scattering pattern quite easily using QuaDRiGa antenna pattern options.
The antenna patterns of the virtual transmitter and receiver's element can be set to $F(\varphi, \theta) = \left( sin(\theta) \right)^{\alpha}$ as in \cite{degli2022reradiation}. Note that the definition of angle $\theta$ in \cite{degli2022reradiation} is different, that is why in our article the antenna pattern includes $sin(\theta)$ instead of $cos(\theta)$. 

In general, due to mutual coupling the center element pattern is different from the edge element pattern \cite{fukao1986numerical}. 
However, for large arrays this effect hardly impacts array gain \cite{fukao1986numerical} and we can assume that all the elements have the same pattern.
Furthermore, mutual coupling causes the difference between the isolated and the embedded element pattern. The embedded pattern can be obtained via EM-simulation or measured in a setup like in \cite{sayanskiy20222d} and uploaded to QuaDRiGa as a custom one. Thus, specifying the pattern of single RIS element is enough, but in this article we used omnidirectional element model for the RIS, BS and the UE so that it would be easier to compare the proposed simulation method with the Rician model.

\section{Numerical results}
\label{sec:Results}

To compare our QuaDRiGa-based RIS channel model against Rician model from \cite{zhou2020joint}, we perform a simulation for a simple single-user, single-RIS scenario.
The BS, the RIS and the UE have coordinates (0, 0, 25), (200, 50, 25) and (250, 0, 1.5) respectively.
The number of elements at the BS, the UE and the RIS is $N_{TX} = 4\times 4 = 16$, $N_{RX} = 4\times1$ and $N_{RIS} = 45 \times 45 = 2025$ respectively. The elements on the RIS, the BS and the UE are $d = \lambda/2$ apart, and for simplicity we model all the elements as isotropic. 
In QuaDRiGa we perform simulation in $1.4$ MHz band with $15$ kHz subcarrier spacing and set the BS power to $30$ dBm, so that the power per subcarrier is $10.3$ dBm.
The noise power spectral density is $-174$ dBm/Hz and the receiver noise factor is $9$ dB. Thus the noise power per subband is $-123$ dBm. 
To compare with Rician model, we extract the channels at one subcarrier and use these values for achievable rate calculation.

In a popular scenario the direct BS-UE path is blocked and the RIS is in Line-of-Sight conditions with respect to the UE and the BS. That is why we choose
3GPP 38.901 UMa NLoS model for $\mathbf{H}_0$ and 3GPP 38.901 UMa LoS model
for $\mathbf{H}_A$ and $\mathbf{H}_B$. 
To make the comparison fair, we set $K = 9$ dB in Rician model \eqref{eq:Rician} for $\mathbf{H}_A$ and $\mathbf{H}_B$, like it is in UMa LoS model. For 
$\mathbf{H}_0$ we set $K = -100$ dB like in UMa NLoS QuaDRiGa configuration file. In \eqref{eq:Rician} we use the same pathloss expressions as in UMa LoS and NLoS models.

Additionally, to determine Large-Scale fading parameters impact, we perform simulation with two modifications of 3GPP 38.901 UMa models. The first one removes the Shadow Fading, setting its standard deviation to $-100$ dB both in LoS and NLoS models.
The second one additionally sets the $K$ standard deviation to $\sigma_K = -100$ dB in UMa LoS model.
We obtain $500$ realizations for every channel and for every model to plot the Cumulative Density Functions (CDFs).

Fig.~\ref{fig:a} shows the CDFs of LoS channels $\mathbf{H}_A$ and $\mathbf{H}_B$. 
Though the average eigenvalue is the same for the Rician (blue lines) and UMa LoS models (red lines), the range of possible eigenvalues is different.
The Rician LoS channel yields eigenvalues in a range of $0.2$ dB, while the eigenvalues in 3GPP UMa LoS channel vary by approximately $20$ dB. Such a large range is caused mainly by the Shadow factor, since setting it to $-100$ dB reduces the eigenvalue range to approximately $3$ dB. Additionally setting $K$ variance to $-100$ dB yields almost the same CDF as with Rician model \eqref{eq:Rician}.

Next, we compare the 3GPP UMa NLoS model against the Rician model from \cite{zhou2020joint}.
As Fig.~\ref{fig:b} demonstrates, the mean values of the eigenvalues differ: with UMa model the first and second eigenvalues are approximately $3$ dB and $7$ dB lower, respectively. 
Moreover, with Rician model the two eigenvalues are only $1.5$ dB apart, while with UMa model the difference between them is $6$ dB.
The reason is that Rician model assumes that the scatterers are distributed uniformly in the hemisphere in front of the receiver, while 3GPP model features specific AoA and AoD ranges with non-uniform distribution that depends on the geometry of the scenario.

Finally, we compare the achievable rate of the RIS-assisted MIMO system for the two models. We assume perfect CSI and use an algorithm suggested by Zhou et al. \cite{zhou2020joint} for joint BS-RIS precoding. As Fig.~\ref{fig:c} shows, there is a huge performance difference between the two models. The mean achievable rate without the RIS is $2.2$ times larger with Rician model. Moreover, with Rician model RIS provides $31 \%$ gain, while with QuaDRiGa RIS gives almost $100 \%$ gain. That is because $\mathbf{H}_0$ eigenvalues are smaller with QuaDRiGa.

To summarize, there is a significant difference in the results obtained with Rician and with 3GPP-standardized models caused by different modeling of the Non-Line-of-Sight part of the channel. Though it is possible to modify Rician model so that the results match for fixed UE, BS and RIS position, the universal calibration is not feasible.

\section{Conclusion}
\label{sec:Conclusion}
Rician channel model, which is extremely popular in RIS analysis, 
is an obstacle to standardization of RIS-assisted MIMO since it models NLoS component in geometry-inconsistent way.
Moreover, Rician model is not compatible with channel model calibration procedure specified in the 3GPP standard. 
To fill this gap, we introduce new RIS-assisted MIMO channel model which is based on 
QuaDRiGa simulation platform with calibrated parameters.
The proposed method inherits realistic 3GPP-compliant channel model from QuaDRiGa.
We demonstrate that, compared with the proposed method, Rician model yields overestimated achievable rate. 
In addition to this, we explain how the scattering pattern of the RIS element can be included in the proposed method.
The channel model we analyzed is difficult to calibrate because it uses isolated RIS element pattern that is different from the embedded one.
In our future research, we are going to obtain the embedded RIS element pattern via measurements and fit the proposed model based on field tests.

\bibliographystyle{IEEEtran}
\bibliography{IEEEabrv,merged_bibliography}

\end{document}